\documentclass[manyauthors,nocleardouble,COMPASS]{cernphprep}

\usepackage{ifpdf}
\ifpdf
  \usepackage{graphicx}
  \usepackage{epstopdf}
\else
  \usepackage{graphicx}
\fi

\usepackage[dvipsnames,usenames]{color}
\usepackage{wrapfig}
\usepackage{amsmath,amssymb}
\usepackage{eurosym}
\usepackage{ulem}
\usepackage{bm}       

\usepackage{siunitx}
\usepackage[english]{babel}

\usepackage{amssymb}
\usepackage{amsfonts}

\usepackage{upgreek}

\input{myunits.sty}

\usepackage{lineno}

\usepackage{fix2col}

\usepackage[font=small,format=plain,labelfont=bf,up]{caption}

\usepackage[T1]{fontenc}
\usepackage{mathrsfs}
\usepackage{rotating}
\usepackage[utf8]{inputenc}
\usepackage{dsfont}
\usepackage[dvipsnames]{xcolor}
\usepackage{pdfpages}
\usepackage{multirow}
\usepackage{array}
\usepackage{hyperref}
\usepackage[percent]{overpic}
\usepackage{accents}
\usepackage{cite}

\usepackage{svn-multi}
\svnid{$Id: g1d_2015.tex 357 2016-11-14 12:11:22Z emk $}

\newcommand{\gp}{g_1^{\rm p}}
\newcommand{\ad}{A_1^{\rm d}}
\newcommand{\gd}{g_1^{\rm d}}

\newcommand{\gn}{g_1^{\rm n}}

\newcommand{\gammad}{\Gamma_1^{\rm d}}

\newcommand{\gammans}{\Gamma_1^{\rm NS}}

\newboolean{bw_mode} 
\setboolean{bw_mode}{false} 
\ifthenelse{
	\boolean{bw_mode}
}{
	\graphicspath{{bw_gfx/}{gfx/}}
}{
	\graphicspath{{gfx/}}
}

\begin{document}
\selectlanguage{english}


\begin{titlepage}
\PHnumber{2016--xxx}
\PHdate{\today}

\vspace{1cm}

\title{Final COMPASS results on the deuteron spin-dependent structure function $\gd$ and the Bjorken sum rule}
%
\Collaboration{The COMPASS Collaboration}
\ShortAuthor{The COMPASS Collaboration}
%
\begin{abstract}
Final results are presented from the inclusive measurement of deep-inelastic polarised-muon 
scattering on longitudinally polarised deuterons using a $^6$LiD target. The data
were taken at 160\,GeV beam energy and the results are shown for the kinematic range 
$1\,(\GeV/c)^2<Q^2<100\,(\GeV/c)^2$ in photon virtuality, $0.004<x<0.7$ in the 
Bjorken scaling variable and $W>4\,\GeV/c^2$ in the mass of the hadronic final state. 
The deuteron double-spin asymmetry 
$\ad$ and the deuteron longitudinal-spin structure function $\gd$ are presented 
in bins of $x$ and $Q^2$. Towards lowest accessible values of $x$, $\gd$ decreases and becomes
consistent with zero within uncertainties. The presented final $\gd$ values together
with the recently published final $\gp$ values of COMPASS are used to again evaluate 
the Bjorken sum rule and perform the QCD fit to the $g_1$ world data
at next-to-leading order of 
the strong coupling constant. In both cases, changes in central values of the 
resulting numbers are well within statistical uncertainties. 
The flavour-singlet axial charge $a_0$, {which is identified in the $\overline{\rm MS}$ renormalisation scheme with the total
contribution of quark helicities to the nucleon spin}, is extracted
from only the COMPASS deuteron data with negligible extrapolation uncertainty: 
$a_0 (Q^2 = 3\,(\GeV/c)^2)=0.32 \pm 0.02_{\rm stat} \pm0.04_{\rm syst} \pm 0.05_{\rm evol}$.
Together with the recent
results on the proton spin structure                                           
function $\gp$,                            
the results on $\gd$ constitute the COMPASS legacy on the  
measurements of $g_1$ through inclusive spin-dependent deep inelastic scattering. 
\end{abstract}

\vspace*{60pt}
Keywords: COMPASS; deep inelastic scattering; spin; structure function; 
parton helicity distributions.\\ \\ \\

\Submitted{to be submitted to Phys. Lett. B}
\vfill
%
%
\section*{The COMPASS Collaboration}
\label{app:collab}
\renewcommand\labelenumi{\textsuperscript{\theenumi}~}
\renewcommand\theenumi{\arabic{enumi}}
\begin{flushleft}
C.~Adolph\Irefn{erlangen},
M.~Aghasyan\Irefn{triest_i},
R.~Akhunzyanov\Irefn{dubna}, 
M.G.~Alexeev\Irefn{turin_u},
G.D.~Alexeev\Irefn{dubna}, 
A.~Amoroso\Irefnn{turin_u}{turin_i},
V.~Andrieux\Irefnn{illinois}{saclay},
N.V.~Anfimov\Irefn{dubna}, 
V.~Anosov\Irefn{dubna}, 
K.~Augsten\Irefnn{dubna}{praguectu}
% ...
\end{flushleft}
%
%
\begin{Authlist}
\item \Idef{turin_p}{University of Eastern Piedmont, 15100 Alessandria, Italy}
\item \Idef{aveiro}{University of Aveiro, Dept.\ of Physics, 3810-193 Aveiro, Portugal}
\item \Idef{bochum}{Universit\"at Bochum, Institut f\"ur Experimentalphysik, 44780 Bochum, Germany\Arefs{l}\Arefs{s}}
\item \Idef{bonniskp}{Universit\"at Bonn, Helmholtz-Institut f\"ur  Strahlen- und Kernphysik, 53115 Bonn, Germany\Arefs{l}}
\item \Idef{bonnpi}{Universit\"at Bonn, Physikalisches Institut, 53115 Bonn, Germany\Arefs{l}}
\item \Idef{brno}{Institute of Scientific Instruments, AS CR, 61264 Brno, Czech Republic\Arefs{m}}
\item \Idef{calcutta}{Matrivani Institute of Experimental Research \& Education, Calcutta-700 030, India\Arefs{n}}
\item \Idef{dubna}{Joint Institute for Nuclear Research, 141980 Dubna, Moscow region, Russia\Arefs{o}}
\item \Idef{erlangen}{Universit\"at Erlangen--N\"urnberg, Physikalisches Institut, 91054 Erlangen, Germany\Arefs{l}}
\item \Idef{freiburg}{Universit\"at Freiburg, Physikalisches Institut, 79104 Freiburg, Germany\Arefs{l}\Arefs{s}}
\item \Idef{cern}{CERN, 1211 Geneva 23, Switzerland}
\item \Idef{liberec}{Technical University in Liberec, 46117 Liberec, Czech Republic\Arefs{m}}
\item \Idef{lisbon}{LIP, 1000-149 Lisbon, Portugal\Arefs{p}}
\item \Idef{mainz}{Universit\"at Mainz, Institut f\"ur Kernphysik, 55099 Mainz, Germany\Arefs{l}}
\item \Idef{miyazaki}{University of Miyazaki, Miyazaki 889-2192, Japan\Arefs{q}}
\item \Idef{moscowlpi}{Lebedev Physical Institute, 119991 Moscow, Russia}
\item \Idef{munichtu}{Technische Universit\"at M\"unchen, Physik Dept., 85748 Garching, Germany\Arefs{l}\Arefs{r}}
\item \Idef{nagoya}{Nagoya University, 464 Nagoya, Japan\Arefs{q}}
\item \Idef{praguecu}{Charles University in Prague, Faculty of Mathematics and Physics, 18000 Prague, Czech Republic\Arefs{m}}
\item \Idef{praguectu}{Czech Technical University in Prague, 16636 Prague, Czech Republic\Arefs{m}}
\item \Idef{protvino}{State Scientific Center Institute for High Energy Physics of National Research Center `Kurchatov Institute', 142281 Protvino, Russia}
\item \Idef{saclay}{IRFU, CEA, Universit\'e Paris-Saclay, 91191 Gif-sur-Yvette, France\Arefs{s}}
\item \Idef{taipei}{Academia Sinica, Institute of Physics, Taipei 11529, Taiwan}
\item \Idef{telaviv}{Tel Aviv University, School of Physics and Astronomy, 69978 Tel Aviv, Israel\Arefs{t}}
\item \Idef{triest_u}{University of Trieste, Dept.\ of Physics, 34127 Trieste, Italy}
\item \Idef{triest_i}{Trieste Section of INFN, 34127 Trieste, Italy}
\item \Idef{triest_ictp}{Abdus Salam ICTP, 34151 Trieste, Italy}
\item \Idef{turin_u}{University of Turin, Dept.\ of Physics, 10125 Turin, Italy}
\item \Idef{turin_i}{Torino Section of INFN, 10125 Turin, Italy}
\item \Idef{illinois}{University of Illinois at Urbana-Champaign, Dept.\ of Physics, Urbana, IL 61801-3080, USA}
\item \Idef{warsaw}{National Centre for Nuclear Research, 00-681 Warsaw, Poland\Arefs{u} }
\item \Idef{warsawu}{University of Warsaw, Faculty of Physics, 02-093 Warsaw, Poland\Arefs{u} }
\item \Idef{warsawtu}{Warsaw University of Technology, Institute of Radioelectronics, 00-665 Warsaw, Poland\Arefs{u} }
\item \Idef{yamagata}{Yamagata University, Yamagata 992-8510, Japan\Arefs{q} }
\end{Authlist}
%
%
\renewcommand\theenumi{\alph{enumi}}
\begin{Authlist}
\item [{\makebox[2mm][l]{\textsuperscript{\#}}}] Corresponding authors
\item [{\makebox[2mm][l]{\textsuperscript{*}}}] Deceased
\item \Adef{a}{Also at Instituto Superior T\'ecnico, Universidade de Lisboa, Lisbon, Portugal}
\item \Adef{b}{Also at Dept.\ of Physics, Pusan National University, Busan 609-735, Republic of Korea and at Physics Dept., Brookhaven National Laboratory, Upton, NY 11973, USA}
\item \Adef{r}{Supported by the DFG cluster of excellence `Origin and Structure of the Universe' (www.universe-cluster.de)}
\item \Adef{d}{Also at Chubu University, Kasugai, Aichi 487-8501, Japan\Arefs{q}}
\item \Adef{x}{Also at Dept.\ of Physics, National Central University, 300 Jhongda Road, Jhongli 32001, Taiwan}
\item \Adef{e}{Also at KEK, 1-1 Oho, Tsukuba, Ibaraki 305-0801, Japan}
\item \Adef{g}{Also at Moscow Institute of Physics and Technology, Moscow Region, 141700, Russia}
\item \Adef{v}{Supported by Presidential grant NSh--999.2014.2}
\item \Adef{h}{Present address: RWTH Aachen University, III.\ Physikalisches Institut, 52056 Aachen, Germany}
\item \Adef{y}{Also at Dept.\ of Physics, National Kaohsiung Normal University, Kaohsiung County 824, Taiwan}
\item \Adef{i}{Present address: Uppsala University, Box 516, 75120 Uppsala, Sweden}
\item \Adef{c}{Supported by the DFG Research Training Group Programmes 1102 and 2044} 
%
%
\item \Adef{l}{Supported by the German Bundesministerium f\"ur Bildung und Forschung}
\item \Adef{s}{Supported by EU FP7 (HadronPhysics3, Grant Agreement number 283286)}
\item \Adef{m}{Supported by Czech Republic MEYS Grant LG13031}
\item \Adef{n}{Supported by SAIL (CSR), Govt.\ of India}
\item \Adef{o}{Supported by CERN-RFBR Grant 12-02-91500}
\item \Adef{p}{\raggedright Supported by the Portuguese FCT - Funda\c{c}\~{a}o para a Ci\^{e}ncia e Tecnologia, COMPETE and QREN,
 Grants CERN/FP 109323/2009, 116376/2010, 123600/2011 and CERN/FIS-NUC/0017/2015}
\item \Adef{q}{Supported by the MEXT and the JSPS under the Grants No.18002006, No.20540299 and No.18540281; Daiko Foundation and Yamada Foundation}
\item \Adef{t}{Supported by the Israel Academy of Sciences and Humanities}
\item \Adef{u}{Supported by the Polish NCN Grant 2015/18/M/ST2/00550}
\end{Authlist}

\end{titlepage}

\begin{flushright}

\end{flushright}
\setpagewiselinenumbers
\section{Introduction}
About a {quarter-century} ago, measurements of the spin-dependent structure
function $g_1^{\rm p}$ by EMC~\cite{emc_spin} at the CERN SPS 
muon beam line led to the very surprising result that, within large experimental 
uncertainties, the quark {spin} contribution to the nucleon spin of 1/2 might be very small 
or even vanishing. This observation initiated enormous experimental and theoretical 
activities aimed at studying the spin structure of the nucleon. In subsequent 
measurements by SMC~\cite{smc}, the same beam line and an upgraded apparatus were 
used to confirm with better precision that only about one third of the spin of the 
nucleon is made up by quark {spins}. This is supported 
by recent lattice QCD simulations \cite{lattice}.

In the last two decades, several new experiments were set up at various laboratories 
to study the longitudinal spin structure of the nucleon in 
{even} more detail, as COMPASS at CERN
using again the same muon beam line at energies $160\,\GeV$ and $200\,\GeV$,
HERMES at DESY using the $27.5\,\GeV$ electron beam of HERA, many experiments at the 
$6\,\GeV$ electron beam of Jefferson Laboratory, as well as PHENIX and STAR at the 
proton-proton collider RHIC with a center of mass energy of $270\,\GeV$. 
Except of the two latter ones, all other experiments studied the 
longitudinal spin structure of the nucleon by inclusive measurements of spin-dependent
deep-inelastic lepton-nucleon scattering (DIS) using longitudinally polarised beams 
and targets, in particular by measuring double-spin cross-section asymmetries. 
More details can be found in recent reviews, see e.g. Ref.~\citen{spin_structur}.

{The measured value of the parton helicity
  contribution to the proton spin is very sensitive to the minimal
  experimental accessible value of the Bjorken-$x$ variable.} Therefore  measurements at low $x$ are crucial
to {understand} the spin structure
{of the nucleon}. According to theoretical 
expectations, new contributions to the DGLAP QCD evolution, e.g.\ double logarithmic terms \cite{double_log},
may be important in this region. Perturbative QCD is considered to be 
applicable for values of $Q^2$ as low as $1\,(\GeV/c)^2$. At COMPASS, using a $160\,\GeV$~muon
beam, this corresponds to a {minimal value of $x$
  equal to} 0.0045. 

In this Letter, results are presented on the longitudinal double-spin asymmetry $\ad$ and 
the longitudinal spin structure function $\gd$ of the deuteron, which are obtained from data 
taken in 2006 with the CERN 160~GeV longitudinally polarised muon beam
and a longitudinally polarised 
$^6$LiD target. The results obtained from the analysis of the 2006 data are described and 
compared to those published earlier~\cite{g1d2006} for the 2002--2004 data. The analysis 
of the combined 2002--2006 data yields the final COMPASS results on $\ad$ and $\gd$.
Moreover, the combined data set analysed in this
work extends to high $Q^2$ values that were {formerly} only reached by SMC, thereby improving
considerably the statisti{cal accuracy}.
{Together with the
results on the proton spin structure
function $\gp$~\cite{g1p2010,compass_2015},
the results for $\gd$ constitute the COMPASS legacy on the
measurements of $g_1$ through inclusive DIS.}

The Letter is organised as follows. Experimental 
set-up and data analysis are described in Sect.~2. The physics context 
of the analysis and details on the calculation of asymmetries are given in Sect.~3. 
In Sect.~4, the results are presented and interpreted. Summary and conclusions are given
in Sect.~5.

\section{Experimental set-up and data analysis}

The COMPASS spectrometer used in 2002--2004 and the upgrades of the polarised
target solenoid and the RICH detector performed in 2005 are described in detail
in Ref.~\citen{nimpaper}.
{In 2006} the target material was $^6$LiD contained in
three cells instead of two.
{They were located along the beam one after the other and} 
had a diameter of {$3\,\Cm$}. The two outermost cells had a length
of $30\,\Cm$ and the central cell was $60\,\Cm$ long. The deuteron polarisation in
$^6$LiD was {$P_{\rm T}\approx 0.52$}, 
{and} the direction of the target polarisation in
the outer cells was opposite to that of the central one. The polarisation direction
was {inverted} on a regular basis by rotating the direction of the {target} solenoid
magnetic field. {Once during the data taking,} the direction of the polarisation with respect
to the solenoid field was {inverted by repolarisation
  in opposite directions keeping the solenoid field unchanged}.
The tertiary M2 beam of the CERN SPS delivered a
naturally polarised muon beam {with a
  polarisation of $P_{\rm B}\approx 0.8$.} The nominal momentum
{was} $160\,\GeV/c$ with a spread
of $5\%$. Momentum and trajectory of each beam particle were measured
by {sets of scintillator hodoscopes},
scintillating fibre and silicon detectors.
The particles produced in an interaction were detected in a two-stage open forward
spectrometer with large momentum and angular acceptance. Each stage contained a
dipole magnet complemented with various tracking detectors {(scintillating
fibre detectors, micropattern gaseous detectors, multiwire proportional chambers,}
drift chambers, straw detectors), as well as hadron and
electromagnetic calorimeters. In the first stage, a RICH detector was used for
hadron identification. Scattered muons were detected by drift tube planes 
{and multiwire proportional chambers} located
behind iron and concrete absorbers. Two types of triggers were used in this analysis.
The ``inclusive'' trigger was based on a signal from a combination of hodoscope
signals from the scattered muon. The ``semi-inclusive'' triggers required an energy
deposition in one of the calorimeters with an optional coincidence with the inclusive
trigger. 

Events with a reconstructed {interaction point} in one of the three target cells are
 selected requiring {at least} a reconstructed incoming muon and a scattered
muon. The measured momentum of the incident muon has to be in  the range
$140\,\GeV/c < p_{\rm B} < 180\,\GeV/c$, and the extrapolated beam track has to cross
 all target cells to equalise the flux through them. The amount of unpolarised
material surrounding the polarised material is minimised by a radial cut on the
 vertex position of $r<1.4\,\Cm$. The scattered muon is identified by requiring
 that it has passed more than $30$ radiation lengths and points to the hodoscope
that triggered the event. In addition, kinematic constraints on the scattering
process are applied. A photon virtuality of $Q^2>1\,(\GeV/c)^2$ is required and
the relative virtual-photon energy has to be in the range $0.1 <y<0.9$. Here,
the lower limit removes events that are difficult to reconstruct, and the upper
limit removes events, the kinematics of which are dominated by radiative effects.
These selection criteria lead to the kinematic range $0.004 < x < 0.7$ and to a
 minimal mass of the hadronic final state of $W>4\,\GeV/c^2$. The final sample
consists of $46$ million events.

\section{Asymmetry calculation}
The longitudinal double-spin asymmetry for one-photon exchange in inclusive DIS on the deuteron, 
$\mu {\rm d} \rightarrow \mu' {\rm X}$, is defined, as function of $x$ and $Q^2$, as 
{
\begin{equation}
\ad = \frac{\sigma_{0}^{\rm T} - \sigma_{2}^{\rm T}}{2 \sigma^{\rm T}},
\end{equation}
with $\sigma_{J}^{\rm T}$ being the $\gamma^*$-deuteron absorption cross section
for total spin projection $J$ in the direction of the virtual photon $\gamma^*$ and 
$\sigma^{\rm T}=(\sigma_{0}^{\rm T}+\sigma_{1}^{\rm T}+\sigma_{2}^{\rm T})/3$ } the deuteron
photoabsorption cross section for transverse virtual photons.
This asymmetry can be derived from the asymmetry 
between the cross sections for parallel and antiparallel oriented longitudinal 
spins of beam particle and target nucleon, where also the 
contribution from the transverse spin asymmetry $A_2^{\rm d}$ has to be taken
into account\footnote{While for a spin-$1/2$ target the first equality in Eq.~(\ref{all}) 
is strict, for a spin-1 target there is an extra contribution in the denominator of 
the asymmetry 
$A_{\rm LL}^{\rm d} = \frac{\sigma^{\uparrow\downarrow}-\sigma^{\uparrow\uparrow}}
{\sigma^{\uparrow\downarrow}+\sigma^{\uparrow\uparrow}+\sigma^{\uparrow 0}}$,
 which is connected to the structure function $b_1$. This function is expected to 
be small~\cite{hoodbhoy}, as also confirmed by a measurement~\cite{hermes-b1}, 
and hence neglected here.}
\begin{equation}
A^{\rm d}_{\rm LL} = \frac                              
{\sigma^{\uparrow \downarrow} - \sigma^{\uparrow \uparrow}}
{\sigma^{\uparrow \downarrow} + \sigma^{\uparrow \uparrow}}
= D (\ad + \eta A_2^{\rm d}).
\label{all}
\end{equation}
Here, the factors
\begin{equation}
\eta = \frac{\gamma(1-y-\gamma^2y^2/4-y^2m^2/Q^2)}{(1+\gamma^2y/2)(1-y/2)-y^2m^2/Q^2}
\end{equation}
and
\begin{equation}
  D = \frac{y ( (1+\gamma^2 y/2)(2-y) - 2y^2 m^2/Q^2 )}{ y^2 (1-2m^2/Q^2) (1+\gamma^2) + 2(1+R) (1-y-\gamma^2 y^2/4)}
\end{equation}
depend only on the kinematics of the process, with $\gamma = 2 M x /\sqrt{Q^2}$;
$m$ and $M$ denote the mass of the muon and the 
nucleon, respectively. The factor $R$ in the depolarisation factor $D$ 
represents the ratio of 
the cross sections for the absorption of a longitudinally and a transversely 
polarised photon by a nucleon. In COMPASS kinematics, the factor $\eta$ and 
the asymmetry $A_2$ are both small, {and hence the contribution $\eta A_2$ is} neglected in the 
calculation of $A_1$ and $g_1$.

For the calculation of the asymmetry, the number of events in each target cell for both polarisation directions can be expressed as
\begin{equation}
N_{i} = a_{i} \phi_{i} n_{i} {\overline \sigma} (1 + P_{\rm B} P_{\rm T} f D \ad)~, ~~~ {i={\rm o1,c1,o2,c2}}~.
\end{equation}
Here, $a_i$ is the acceptance, $\phi_i$ the incoming flux, $n_i$ the number of
target nucleons, {${\overline \sigma}$} 
the spin-averaged cross section and $f$ the dilution factor.
{There are four equations} {describing the two
solenoid field directions (1,2) for the combined outer cells (o) and
the central cell (c).}
They are combined into one second-order equation in $A_1$ for the ratio 
$(N_{\rm o1} N_{\rm c2})/(N_{\rm o2} N_{\rm c1})$, where acceptance and flux cancel. The 
asymmetry is calculated for periods of stable data taking, which are combined 
using the weighted mean. In order to minimise the statistical uncertainty, in 
the asymmetry calculation each event is used with a weight factor
\begin{equation}
w = P_{\rm B} f D.
\end{equation}

Systematic uncertainties are calculated taking into account multiplicative 
and additive contributions. The multiplicative contribution {$\Delta A_1^{\rm mult}$}
{comprises} the uncertainties on beam and target polarisations and the 
uncertainties on depolarisation and dilution factors. The size of each of these 
contributions is shown in Table~\ref{tab:syst}. It also shows the additive 
contributions {from i)~possible false asymmetries, 
ii)~the neglect of the transverse asymmetry $A_2$ and iii)~the
  uncertainty on spin-dependent radiative corrections. False asymmetries
  are investigated using two methods.}
In one method, possible false asymmetries are studied by calculating 
the asymmetry between cells with the same polarisation direction, i.e.\ between 
both outermost target cells and for the two halves of the central cell. Both 
asymmetries are found to be consistent with zero. In the other method, 
``pulls`` \cite{compass_npb765} are used to check for time-dependent effects. 
Here, the asymmetry is calculated for each subsample and compared to the final 
asymmetry. No significant broadening is observed in these distributions. The 
statistical limitation of this method leads to an uncertainty between $38\%$ 
and $75\%$ of the statistical uncertainty, {which
  represents the largest additive contribution.}

\begin{table}[!htbp]
	\centering
	\caption{Summary for the systematic uncertainty of $A_1$.}
	\label{tab:syst}
	\begin{tabular}{|ll|c|}
		\hline
		Beam polarisation		& $\Delta P_{\rm B}/P_{\rm B}$			& $5\%$ \\
		Target polarisation		& $\Delta P_{\rm T}/P_{\rm T}$			& $5\%$ \\
		Depolarisation factor	& $\Delta D(R)/D(R)$		& $2 - 3\%$ \\
		Dilution factor			& $\Delta f/f$ 				& $2 - 3\%$ \\
		Total					& $\Delta A_1^{\rm mult}$	& $\simeq 0.08\cdot \ad$ \\
		\hline
		\hline
		False asymmetry			& $A_{\rm false}$					& $< 0.75 \cdot \Delta A_1^{\rm stat}$ \\
		Transverse asymmetry	& $\eta \cdot A_2^{\rm d}$	& $< 10^{-4}$\\
		Rad. corrections		& $A^{\rm RC}$				& $10^{-5} - 10^{-3}$ \\

		\hline
	\end{tabular}
\end{table}
\section{Results}
\begin{figure}[htbp]
	\centering
	\includegraphics[width=0.9\textwidth]{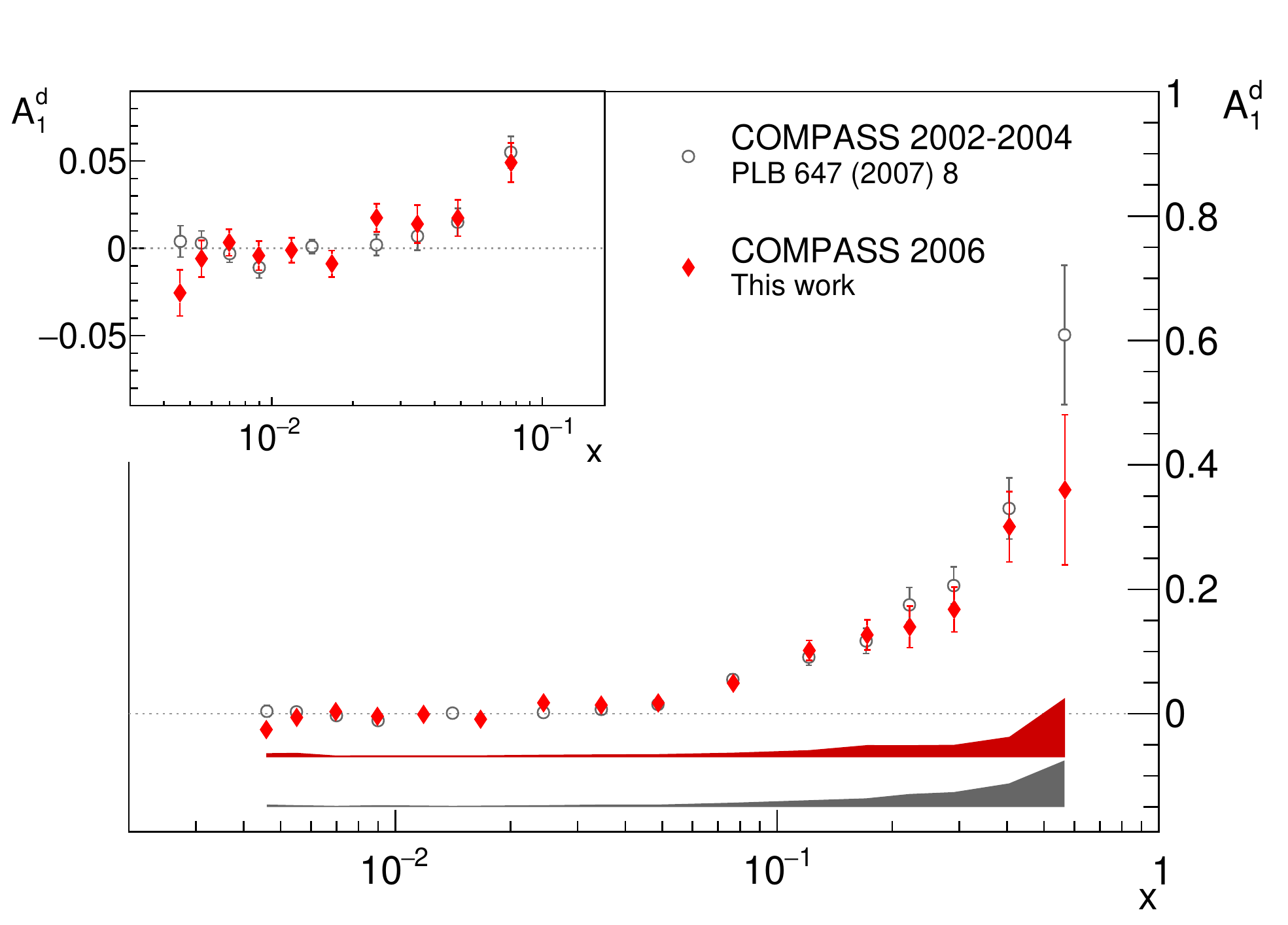}
	\caption{Comparison between the results on $\ad$ obtained from the 2006 data set and the previous results from COMPASS.}
	\label{fig:comp_04_06}
\end{figure}

\begin{figure}[htbp]
	\centering
	\includegraphics[width=0.9\textwidth]{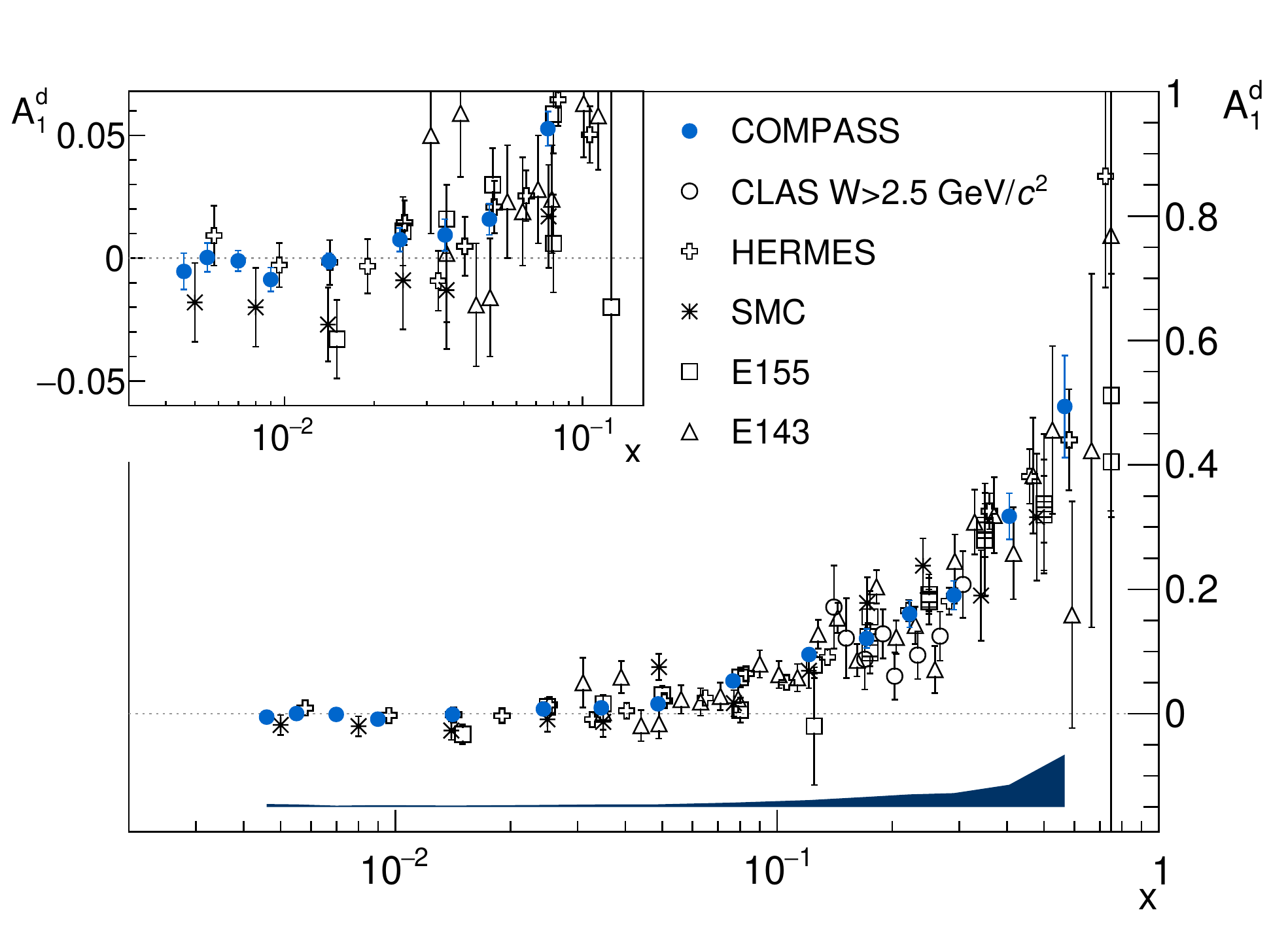}
	\caption{Comparison between the combined COMPASS results on $\ad$ {and the} world data (CLAS \cite{CLAS}, 
        HERMES \cite{Hermes}, SMC \cite{smc}, E155 \cite{e155d} and E143 \cite{e143}). 
        {All data points are shown at their measured $Q^2$ values.}}
	\label{fig:a1d_world}
\end{figure}

\begin{figure}[!htbp]
	\centering
	\includegraphics[width=0.85\textwidth]{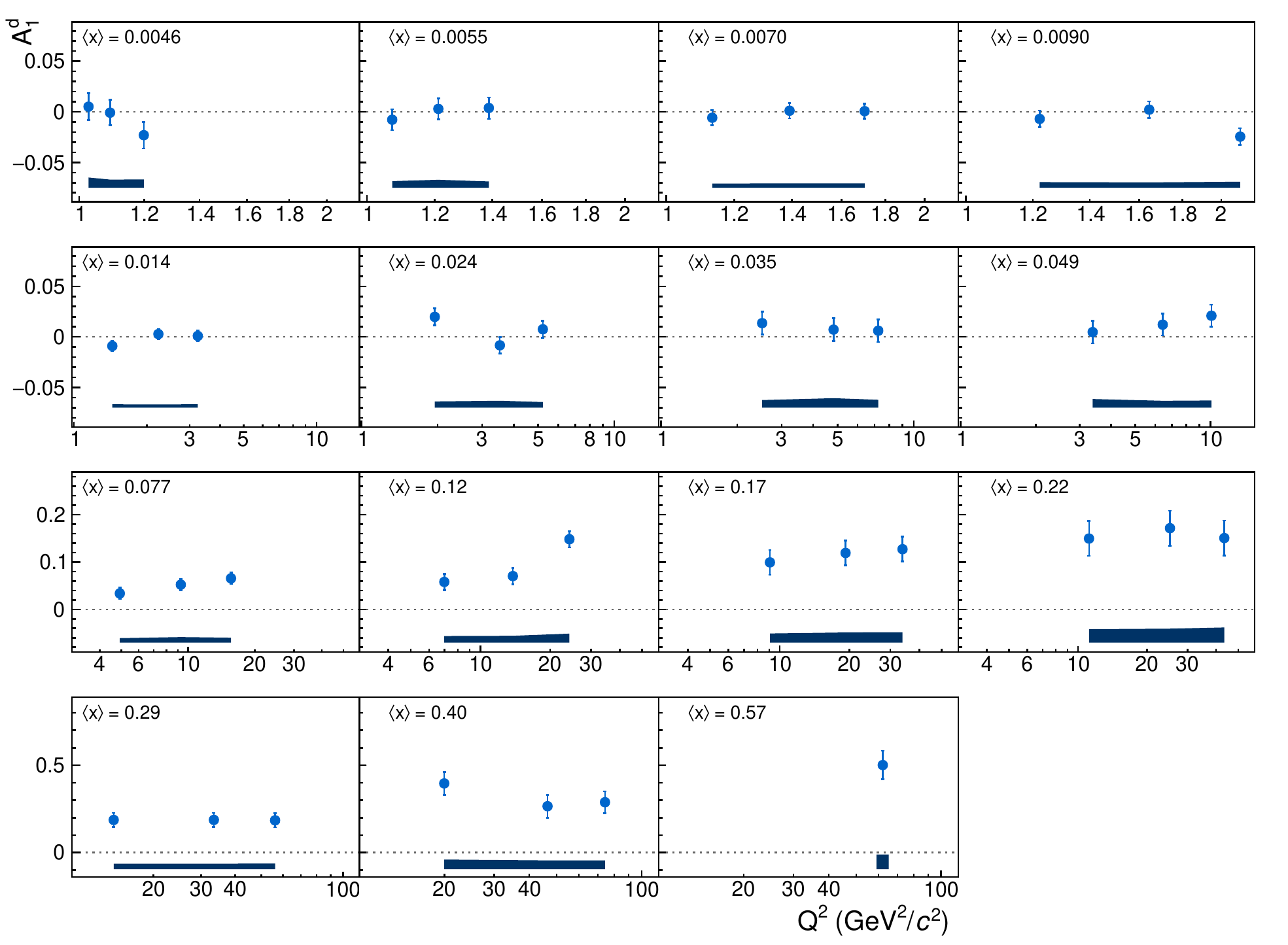}
	\caption{Results on $\ad$ from the combined COMPASS data in bins of $x$ and $Q^2$.}
	\label{fig:a1_x_q2}
\end{figure}

The double-spin asymmetry $\ad$ and the spin-dependent structure function $\gd$
are calculated in bins of $x$~and~$Q^2$. In Figure~\ref{fig:comp_04_06}, the 
results in bins of $x$ obtained from the 2006 data set are compared to the 
results from the 2002--2004 data~\cite{g1d2006},
{which demonstrates} the good agreement between both data sets 
(the $\chi^2$ probability is $63\%$).
The 2006 data {increase} the statistics
of the 2002--2004 data {by approximately} {50\%}. The results from both
data sets are combined using the weighted mean. In
Fig.~\ref{fig:a1d_world}, the combined COMPASS results 
on $\ad$ are compared to the world data on 
$\ad$ at the measured {values of} $Q^2$. 
All data sets agree well with one another. The data confirm the well-known weak $Q^2$ 
dependence of the asymmetry.
This is also illustrated in Fig.~\ref{fig:a1_x_q2} that shows the
$Q^2$ dependence of the COMPASS data for each $x$ bin. No
{clear} dependence on {$Q^2$} is visible 
in any $x$ bin. 
The numerical values of the combined data for $\ad(x)$ and $\ad(x,Q^2)$ are given
in Appendix A in Tables~\ref{tab:a1_x_02_06} and \ref{tab:a1_x_q2_02_06_2}. 


\begin{figure}[!htbp]
	\centering
	\includegraphics[width=0.90\textwidth]{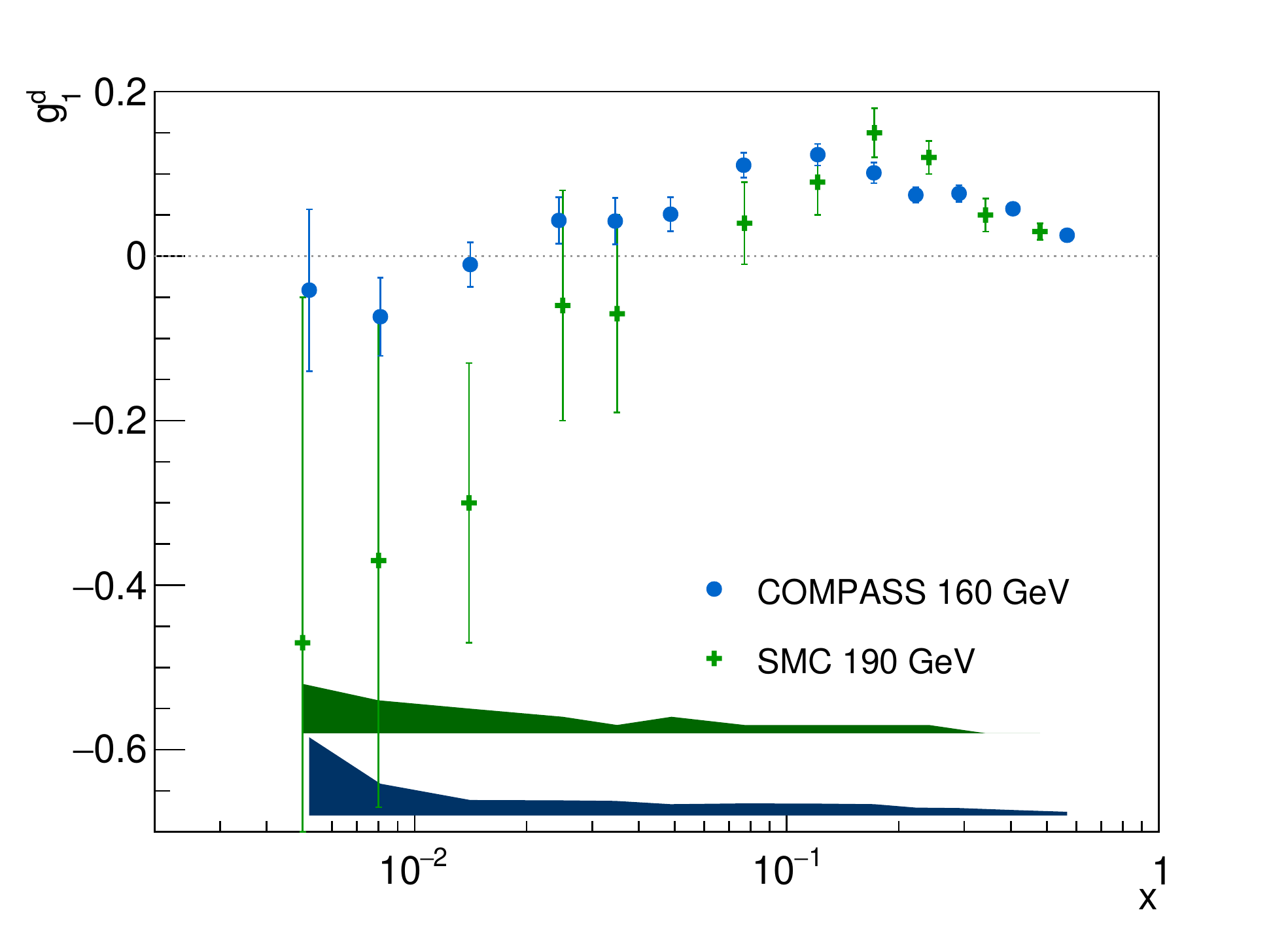}
	\caption{Comparison between SMC \cite{smc} and combined COMPASS results on $\gd$. The 
        systematic uncertainty is illustrated by the bands at the bottom. 
        {All data points are shown at their measured $Q^2$ values.}}
	\label{fig:g1_smc}
\end{figure}

The spin-dependent structure function $\gd$ is calculated from the asymmetry $\ad$ using
\begin{equation} 
 \gd(x,Q^2) = \frac{F_2^{\rm d}(x,Q^2)}{2x(1+R(x,Q^2))} A_1^{\rm d}(x,Q^2)\,.
\end{equation}
The parametrisation of the unpolarised structure
function $F_2^{\rm d}$ is  taken from Ref.~\citen{smc} and the parametrisation of the ratio $R$ is 
taken from Ref.~\citen{r1998}. The $x$ dependence of the structure function is shown in 
Fig.~\ref{fig:g1_smc} together with the results from SMC \cite{smc} that were obtained at a 
higher beam energy of 190~GeV. In the figure, the two
COMPASS data points at lowest $x$ are obtained as averages from the four
lowest $x$ bins {used in this} 
analysis. The systematic uncertainties are shown by bands at the bottom. 
The COMPASS data do not support large negative values of the structure
function at low $x$, {an indication of which may be seen in the SMC data}. 
Instead, $g_1^{\rm d}$ is compatible with zero for $x$ decreasing towards the lower 
limit of the measured range. 

\begin{figure}[!htbp]
	\centering
	\includegraphics[width=0.76\textwidth]{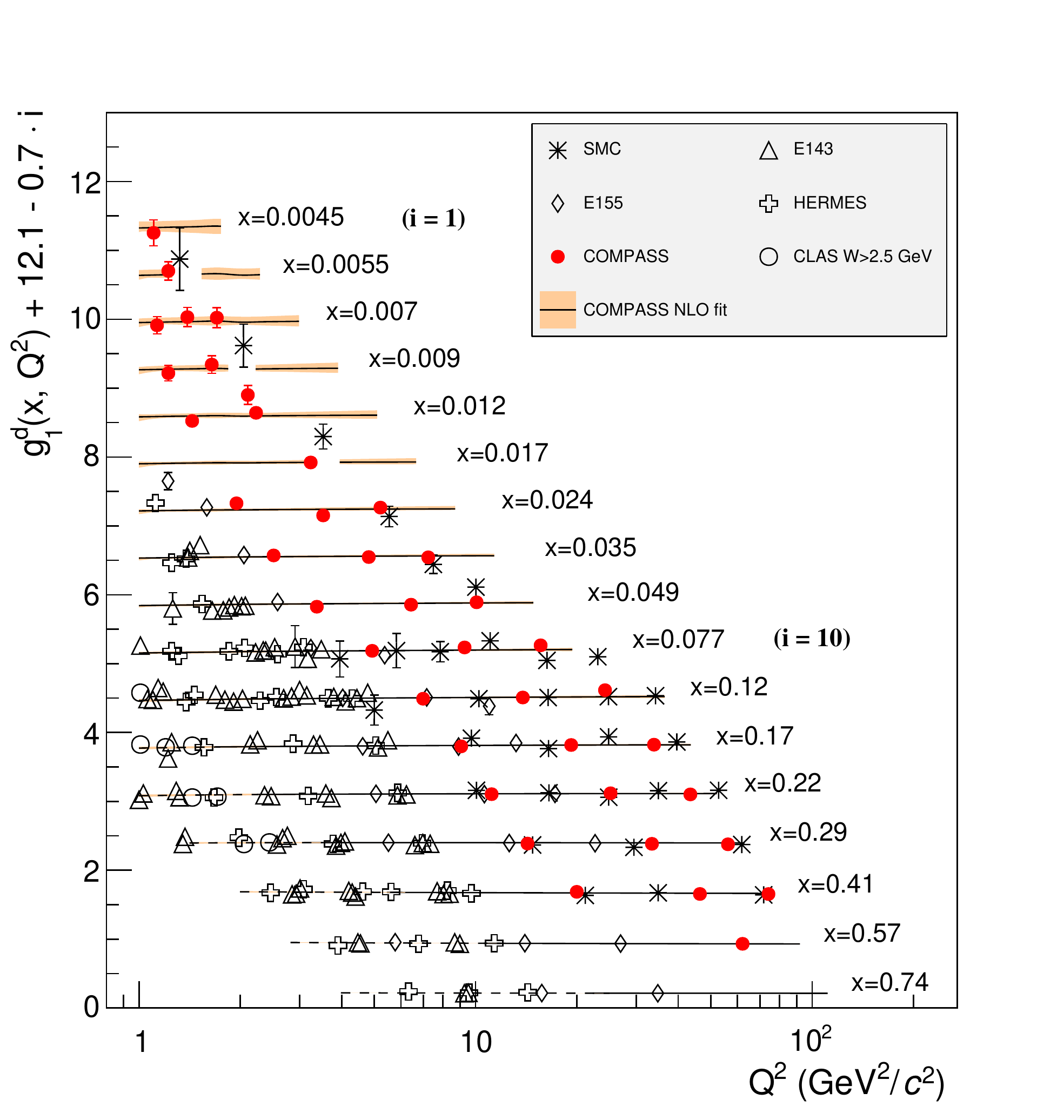}
	\caption{World data on the spin-dependent structure function $\gd$ as a function of 
$Q^2$ for various values of $x$ with the combined COMPASS data as filled circles. The lines 
represent the $Q^2$ dependence for each value of $x$ as determined from the updated NLO QCD 
fit {to the world data}. The dashed parts represent the region with $W^2 < 10\,(\GeV/c^2)^2$.}
	\label{fig:g1d_x_q2}
\end{figure}

{The new results on the spin-dependent structure
  function $g_1^{\rm d}$, which are shown in Fig.~\ref{fig:g1d_x_q2}
  together with the world data in bins of $x$ and $Q^2$, constitute the final
COMPASS results and hence supersede the ones published              
in Ref.~\citen{g1d2006}. They improve the statistical 
precision of the combined world data on $\gd$, in particular at low $x$ where
SMC is the only other experiment that contributes.}

The NLO QCD fit on the $g_1$ world data described in detail in 
Ref.~\citen{compass_2015} is 
repeated using the updated results for $\gd$. 
The fit results are shown as curves in 
Fig.~\ref{fig:g1d_x_q2} for the various $x$ bins. Compared to the previous analysis, the changes 
in central values of resulting parameters are of the order of statistical uncertainties.
The parameters of the QCD fit are available together with the deuteron results on 
HepData~\cite{hepdata}.

The presented final $\gd$ values together with the final 
COMPASS results on $\gp$~\cite{g1p2010,compass_2015} 
are used to re-evaluate the Bjorken sum rule as described in the 
same reference. The results
\begin{equation}
\gammans = 0.192 \pm 0.007_{\rm stat} \pm 0.015_{\rm syst}~~~~~
{{\rm and}}~~~~~~|g_{_{\rm A}}/g_{_{\rm V}}| = 1.29 \pm 0.05_{\rm stat} \pm 0.10_{\rm syst}
\end{equation}
agree within statistical errors with the previously published ones.

The new combined data are also used to update the results for the first moment
of the spin-dependent structure function of the nucleon, 
$\Gamma_1^{\rm N}(Q^2)=\int_0^1 \gd(x,Q^2)/(1-1.5\omega_D) \text{d}x$,
where  
{$\omega_D=-0.05 \pm 0.01$~\cite{machleidt}} is the correction
for the D-state admixture in the deuteron. 
The first moment is calculated by evolving the values of $\gd$ to the
common {value} $Q^2 = 3\,(\GeV/c)^2$. From these values the contribution to the first moment 
from the measured $x$ range is calculated. In order to evaluate the contributions from the 
unmeasured regions, an extrapolation from the QCD fit to $x=0$ and $x=1$ is used. The updated 
value of the first moment from COMPASS data alone is:

\begin{equation}
	\Gamma_1^{\rm N}(Q^2 = 3\,(\GeV/c)^2) = 0.046 \pm 0.002_{\rm stat} \pm 0.004_{\rm syst} \pm 0.005_{\rm evol}\,.
\end{equation}

The systematic uncertainty is dominated
by bin-to-bin correlated uncertainties of $P_B, P_T, f, D$ and $F_2$,
while the impact of possible false asymmetries largely cancels
in the discussed integral.
The contributions from the different $x$ ranges are shown in Table~\ref{tab:first_mom}. 
The contributions from both extrapolation regions are {very small 
and their uncertainties negligible}. 

\begin{table}[!h]
	\caption{Contributions to the first moment of $g_1^{\rm N}$ at $Q^2 =3\,(\GeV/c)^2$ with statistical uncertainties from the COMPASS data.}
	\label{tab:first_mom}
	\centering
	\begin{tabular}{|rcl|c|}
		\hline
		\multicolumn{3}{|c|}{$x$ range} & $\Gamma_1^{\rm N}$\\
		\hline
		$0$  	&$\!\!\!-\!\!\!$& $0.004$	 & $-0.001$\\
		$0.004$	&$\!\!\!-\!\!\!$& $0.7$	 & $0.045 \pm 0.002$ \\ 
		$0.7$  	&$\!\!\!-\!\!\!$& $1.0$	 & $0.001$\\
		\hline
	\end{tabular}
\end{table}

\begin{figure}[!htbp]
	\centering
	\includegraphics[width=0.9\textwidth]{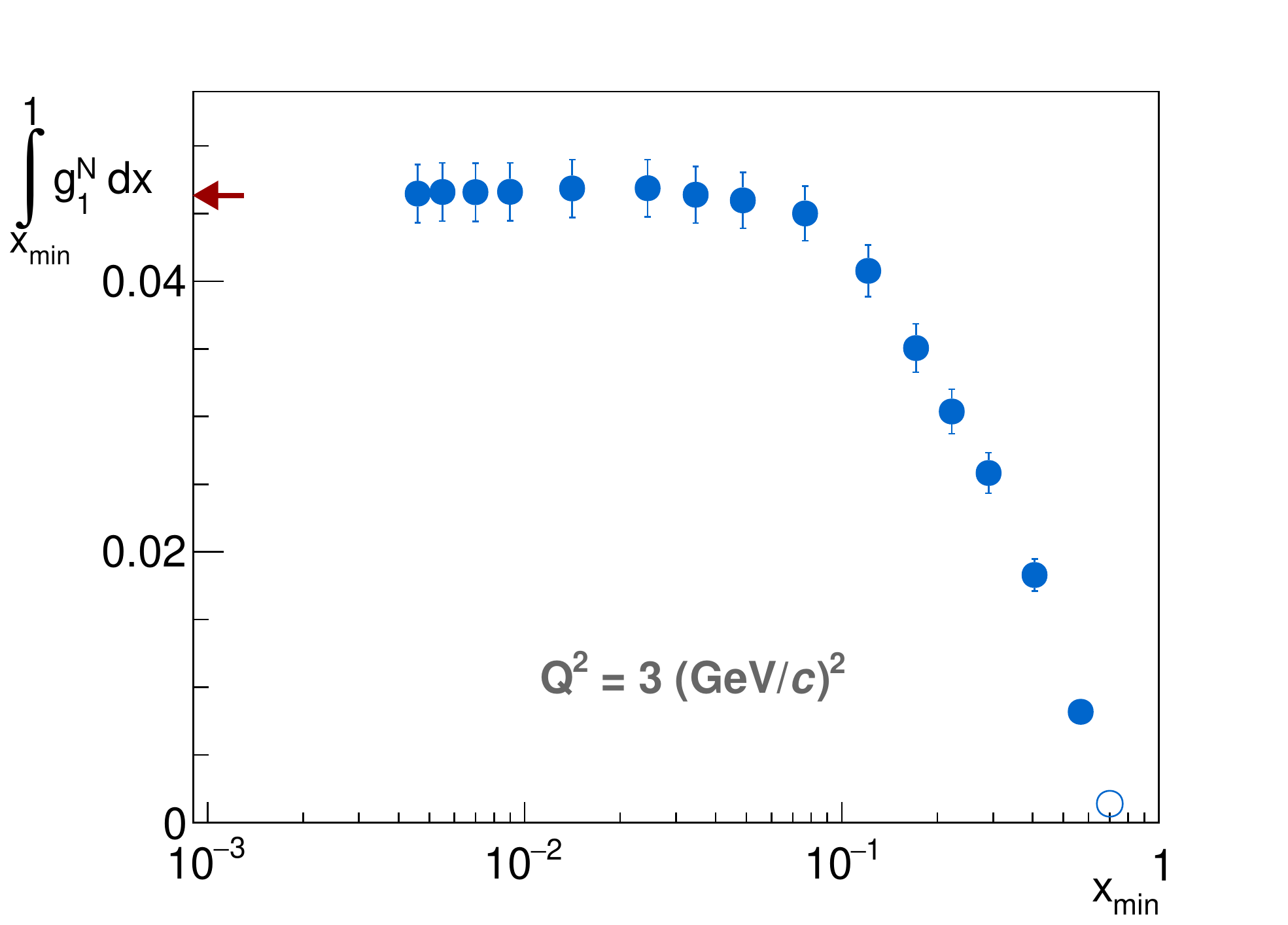}
	\caption{Values of $\int_{x_{\rm min}}^1 \gd/(1-1.5\omega_D) dx$ as a function of $x_{\rm min}$. The open circle at $x = 0.7$ is obtained from the fit. The arrow on the left side shows the value for the full range, $0 \leq x \leq 1$.}
	\label{fig:g1d_mom}
\end{figure}

All presently available experimental information supports the observation that $\gd$ vanishes
when $x$ decreases down to the lowest accessible values. As can be seen in 
Fig.~\ref{fig:g1d_mom}, the first moment of $\gd$ measured from only the 
COMPASS deuteron data approaches its asymptotic value already in the 
experimentally accessible region for $Q^2=3\,(\GeV/c)^2$. It can 
hence be used for physics interpretation without using 
proton data and without invoking the Bjorken sum rule.  

The structure function $\gd$ as physical observable is
factorisation-scheme independent, whereas its representation as convolution(s)
of quark, anti-quark, and gluon helicity distributions with respective Wilson 
coefficient functions~\cite{structureconv,splitting} involves a possible 
scheme dependence. In the `modified minimal  subtraction' ($\overline{\rm MS}$)
factorisation scheme~\cite{msbar}, the first moment 
of the gluon coefficient function vanishes, and hence the first moment 
$\gammad$ does not depend on the gluon helicity distribution. This
allows {for} the direct determination of the 
flavour-singlet axial charge $a_0$ from the COMPASS $\gammad$ result using 
only the axial charge $a_8$ as additional input:

\begin{equation}
 a_0(Q^2) = \frac{1}{\Delta C^{\overline{\rm MS}}_{S} (Q^2)} \left[ \frac{9 \gammad}{(1-1.5\omega_D)} - \frac{1}{4} a_8  \Delta C^{\overline{\rm MS}}_{NS}(Q^2) \right]
\end{equation}
Here, $\Delta C^{\overline{\rm MS}}_{S}(Q^2)$ and $\Delta C^{\overline{\rm MS}}_{NS}(Q^2)$ are the singlet and non-singlet coefficient functions, which are calculated up to the third order in $\alpha_S(Q^2)$ in perturbative QCD in Ref.~\citen{larin}.
In the $\overline{\rm MS}$ factorisation scheme, $a_0$ is identified with the total quark contribution to the nucleon spin:
$a_0 \overset{{\overline{\rm MS}}}{=} \Delta \Sigma = (\Delta u + \Delta \bar{u}) + (\Delta d + \Delta \bar{d}) + (\Delta s + \Delta \bar{s})$. Here, $\Delta \accentset{(-)}{f}$ is the helicity distribution of flavour-$f$ quarks integrated over the measured $x$-range.

The integral $\gammad$ is calculated directly using the measured combined 
COMPASS deuteron data points. The evolution of a given data point to the 
common value $Q^2 = 3\,(\GeV/c)^2$ is obtained as average from the variety 
of the updated NLO QCD fits along the lines described in 
Ref.~\citen{compass_2015}, whereby the range in fit solutions is transformed 
into an evolution uncertainty for this 
point. With $\alpha_s = 0.337 \pm 0.012$ for $Q^2 =3\,(\GeV/c)^2$ and the {corresponding}
values for $\Delta C_{\overline{\rm MS}}^{S}(Q^2)$ and 
$\Delta C_{\overline{\rm MS}}^{NS}(Q^2)$ to 
${\cal O}({\alpha_{s}})$, together with 
$a_8 = 0.585 \pm 0.025$ \cite{axial}, the flavour-singlet axial charge is obtained as 
\begin{equation}
	a_0(Q^2 = 3\,(\GeV/c)^2) = 0.32 \pm 0.02_{\rm stat} \pm 0.04_{\rm syst} \pm 0.05_{\rm evol}\,.
	\label{eq:a0}
\end{equation}
In the context of the above discussed asymptotic behaviour of $\gammad$, no 
additional extrapolation uncertainties occur. 
The largest contribution to the {total uncertainty
originates} from the uncertainties in the 
evolution to a common $Q^2$, the reason behind being the large uncertainty of 
the polarised gluon distribution {obtained} in the fits.
This independent result on $a_0$ is consistent with the value of $a_0$ 
obtained from the COMPASS {NLO} QCD fit \cite{compass_2015} of the world data. 
Note the remarkably good statistical and systematic accuracy of this result
obtained from only the COMPASS deuteron data when comparing to the corresponding
accuracy of the fit result \cite{compass_2015}.

\section{Summary and conclusions}
We have presented new results on the longitudinal spin structure function $\gd$ from data 
taken in 2006 and we have combined them with our previously measured ones. All data were 
taken using the $160\,\GeV$ CERN muon beam and a longitudinally polarised $^{6}$LiD target. 
The results cover the kinematic range  $0.004< x <0.7$\,, $1\,(\GeV/c)^2 < Q^2 < 100\,(\GeV/c)^2$ and $W>4\,\GeV/c^2$.
The double-spin asymmetry is studied in bins of $x$ and $Q^2$. The combined results for 
$\gd$ at low $x$ $(x< 0.03)$ improve
{considerably} the precision compared to the only existing 
result in this region, 
which originates from SMC, {so that $\gd$ appears now compatible with zero
at the presently lowest accessible values of $x$.} 
The combined set of data was included in our NLO QCD fit to the $\gp$, $\gd$ and $\gn$ 
world data. In addition,
a re-evaluation of the Bjorken sum rule was performed using only COMPASS results.
Both, for the QCD {NLO} fit and the Bjorken sum rule,
the new values stay compatible with the published ones within statistical 
uncertainties. Finally, the COMPASS deuteron data alone lead to a determination of the 
flavour-singlet axial charge 
$a_0 = 0.32 \pm 0.02_{\rm stat} \pm 0.04_{\rm syst} \pm 0.05_{\rm evol}$ at 
$Q^2 = 3\,(\GeV/c)^2$ from the first moment of $\gd$ with negligible
extrapolation {uncertainty}.
Together with the                  
results on the proton spin structure                                           
function $\gp$~\cite{g1p2010,compass_2015},                            
the results for $\gd$ constitute the COMPASS legacy on the  
measurements of $g_1$.

\section*{Acknowledgements}
We gratefully acknowledge the support of the CERN management and staff and the
skill and effort of the technicians of our collaborating institutes.  This work
was made possible by the financial support of our funding agencies.  

\clearpage
\appendix 
{\setlength{\textheight}{53\baselineskip}
\section{Appendix}
 
The results for $A_1^{\rm d}$ and $g_1^{\rm d}$  are given
in Tables~\ref{tab:a1_x_02_06} and \ref{tab:a1_x_q2_02_06_2}.

\begin{table}[!h]
\centering
\caption{Values for $\ad$ and $\gd$ as a function of $x$ at the
measured values of $Q^2$ for the
combined 2002--2006 data. The first uncertainty is statistical, the
second one is systematic.}
\label{tab:a1_x_02_06}
\footnotesize
\begin{tabular}{|rcl|S[table-format=1.4]|S[table-format=2.2]|
                 S[table-format=2.4]@{\,$\pm$\,}S[table-format=1.4]@{\,$\pm$\,}S[table-format=1.4]|
                 S[table-format=2.4]@{\,$\pm$\,}S[table-format=1.4]@{\,$\pm$\,}S[table-format=1.4]|}
\hline
\multicolumn{3}{|c|}{$x$ range} & \multicolumn{1}{c|}{$\langle x \rangle$} & \multicolumn{1}{c|}{$\langle Q^2 \rangle
((\GeV/c)^2) $} & \multicolumn{3}{c|}{$\ad$} & \multicolumn{3}{c|}{$\gd$} \\
\hline
\hline
0.004 & \!\!\!--\!\!\!&      0.005 & 0.0046      & 1.10
& -0.0054 & 0.0074 & 0.0048       & -0.13   & 0.17   &0.11 \\
0.005 & \!\!\!--\!\!\!&      0.006 & 0.0055      & 1.22
&  0.0003 & 0.0058 & 0.0043       &  0.00   & 0.12   &0.09 \\
0.006 & \!\!\!--\!\!\!&      0.008 & 0.0070      & 1.39
& -0.0011 & 0.0042 & 0.0023       & -0.016  & 0.071  &0.040 \\
0.008 & \!\!\!--\!\!\!&      0.010 & 0.0090      & 1.62
& -0.0087 & 0.0049 & 0.0031       & -0.121  & 0.064  &0.038 \\
0.010 & \!\!\!--\!\!\!&      0.020 & 0.0141      & 2.19
& -0.0011 & 0.0032 & 0.0024       & -0.010  & 0.027  &0.019 \\
0.020 & \!\!\!--\!\!\!&      0.030 & 0.0244      & 3.29
&  0.0075 & 0.0048 & 0.0034       &  0.043  & 0.028  &0.018 \\
0.030 & \!\!\!--\!\!\!&      0.040 & 0.0346      & 4.43
&  0.0095 & 0.0064 & 0.0042       &  0.043  & 0.028  &0.018 \\
0.040 & \!\!\!--\!\!\!&      0.060 & 0.0487      & 6.06
&  0.0159 & 0.0063 & 0.0044       &  0.051  & 0.021  &0.014 \\
0.060 & \!\!\!--\!\!\!&      0.100 & 0.0766      & 9.00
&  0.0527 & 0.0070 & 0.0072       &  0.111  & 0.015  &0.015 \\
0.100 & \!\!\!--\!\!\!&      0.150 & 0.121       & 13.5
&  0.095  & 0.010  & 0.011        &  0.123  & 0.013  &0.014 \\
0.150 & \!\!\!--\!\!\!&      0.200 & 0.171       & 18.6
&  0.121  & 0.015  & 0.016        &  0.101  & 0.013  &0.014 \\
0.200 & \!\!\!--\!\!\!&      0.250 & 0.222       & 23.8
&  0.160  & 0.022  & 0.020        &  0.0744 & 0.0096 &0.0096 \\
0.250 & \!\!\!--\!\!\!&      0.350 & 0.290       & 31.1
&  0.190  & 0.023  & 0.022        &  0.076  & 0.010  &0.009 \\
0.350 & \!\!\!--\!\!\!&      0.500 & 0.405       & 43.9
&  0.317  & 0.037  & 0.036        &  0.0576 & 0.0069 &0.0067 \\
0.500 & \!\!\!--\!\!\!&      0.700 & 0.567       & 60.8
&  0.494  & 0.082  & 0.084        &  0.0254 & 0.0042 &0.0045 \\
\hline
\end{tabular}
\end{table}

\begin{table}[!h]
\centering
\caption{Values for $\ad$ and $\gd$ as a function of $x$ and $Q^2$ for the combined 2002--2006 data. The first uncertainty is statistical, the second one is systematic.}
\label{tab:a1_x_q2_02_06_2}
\footnotesize
\begin{tabular}{|rcl|S[table-format=1.4]|S[table-format=2.2]|
                 S[table-format=2.4]@{\,$\pm$\,}S[table-format=1.4]@{\,$\pm$\,}S[table-format=1.4]|
                 S[table-format=2.4]@{\,$\pm$\,}S[table-format=1.4]@{\,$\pm$\,}S[table-format=1.4]|}
\hline
\multicolumn{3}{|c|}{$x$ range} & \multicolumn{1}{c|}{$\langle x \rangle$} & \multicolumn{1}{c|}{$\langle Q^2 \rangle
((\GeV/c)^2) $} & \multicolumn{3}{c|}{$\ad$} & \multicolumn{3}{c|}{$\gd$} \\
\hline
\hline
0.004	& \!\!\!--\!\!\!& 0.005	& 0.0045 & 1.03	&  0.005  & 0.013  & 0.010	&  0.12   & 0.30   & 0.23 \\
0.004	& \!\!\!--\!\!\!& 0.005	& 0.0046 & 1.09	& -0.001  & 0.013  & 0.008	& -0.02   & 0.29   & 0.19 \\
0.004	& \!\!\!--\!\!\!& 0.005	& 0.0047 & 1.20	& -0.023  & 0.013  & 0.008	& -0.54   & 0.30   & 0.19 \\
\hline
0.005	& \!\!\!--\!\!\!& 0.006	& 0.0055 & 1.07	& -0.008  & 0.010  & 0.007	& -0.15   & 0.20   & 0.12 \\
0.005	& \!\!\!--\!\!\!& 0.006	& 0.0055 & 1.21	&  0.003  & 0.010  & 0.008	&  0.06   & 0.21   & 0.16 \\
0.005	& \!\!\!--\!\!\!& 0.006	& 0.0056 & 1.39	&  0.004  & 0.011  & 0.006	&  0.08   & 0.22   & 0.14 \\
\hline
0.006	& \!\!\!--\!\!\!& 0.008	& 0.0069 & 1.13	& -0.0058 & 0.0075 & 0.0042	& -0.09   & 0.11   & 0.06 \\
0.006	& \!\!\!--\!\!\!& 0.008	& 0.0069 & 1.39	&  0.0011 & 0.0075 & 0.0043	&  0.02   & 0.12   & 0.07 \\
0.006	& \!\!\!--\!\!\!& 0.008	& 0.0072 & 1.70	&  0.0007 & 0.0075 & 0.0043	&  0.01   & 0.13   & 0.07 \\
\hline
0.008	& \!\!\!--\!\!\!& 0.010	& 0.0089 & 1.22	& -0.0070 & 0.0084 & 0.0055	& -0.08   & 0.10   & 0.07 \\
0.008	& \!\!\!--\!\!\!& 0.010	& 0.0089 & 1.65	&  0.0021 & 0.0083 & 0.0052	&  0.03   & 0.11   & 0.07 \\
0.008	& \!\!\!--\!\!\!& 0.010	& 0.0091 & 2.11	& -0.0245 & 0.0083 & 0.0059	& -0.36   & 0.12   & 0.09 \\
\hline
0.010	& \!\!\!--\!\!\!& 0.020	& 0.0132 & 1.44	& -0.0090 & 0.0051 & 0.0034	& -0.076  & 0.043  & 0.029 \\
0.010	& \!\!\!--\!\!\!& 0.020	& 0.0135 & 2.23	&  0.0028 & 0.0051 & 0.0033	&  0.027  & 0.050  & 0.032 \\
0.010	& \!\!\!--\!\!\!& 0.020	& 0.0156 & 3.24	&  0.0009 & 0.0051 & 0.0034	&  0.009  & 0.049  & 0.033 \\
\hline
0.020	& \!\!\!--\!\!\!& 0.030	& 0.0239 & 1.95	&  0.0198 & 0.0082 & 0.0062	&  0.101  & 0.042  & 0.032 \\
0.020	& \!\!\!--\!\!\!& 0.030	& 0.0240 & 3.53	& -0.0083 & 0.0082 & 0.0069	& -0.051  & 0.050  & 0.042 \\
0.020	& \!\!\!--\!\!\!& 0.030	& 0.0253 & 5.22	&  0.0075 & 0.0082 & 0.0056	&  0.048  & 0.053  & 0.037 \\
\hline
0.030	& \!\!\!--\!\!\!& 0.040	& 0.0342 & 2.51	&  0.014  & 0.011  & 0.008	&  0.052  & 0.043  & 0.029 \\
0.030	& \!\!\!--\!\!\!& 0.040	& 0.0344 & 4.82	&  0.007  & 0.011  & 0.009	&  0.033  & 0.051  & 0.043 \\
0.030	& \!\!\!--\!\!\!& 0.040	& 0.0352 & 7.24	&  0.006  & 0.011  & 0.008	&  0.029  & 0.054  & 0.038 \\
\hline
0.040	& \!\!\!--\!\!\!& 0.060	& 0.0477 & 3.38	&  0.005  & 0.011  & 0.009	&  0.014  & 0.032  & 0.025 \\
0.040	& \!\!\!--\!\!\!& 0.060	& 0.0482 & 6.43	&  0.012  & 0.011  & 0.007	&  0.040  & 0.036  & 0.023 \\
0.040	& \!\!\!--\!\!\!& 0.060	& 0.0502 & 10.1	&  0.021  & 0.011  & 0.007	&  0.072  & 0.037  & 0.025 \\
\hline
0.060	& \!\!\!--\!\!\!& 0.100	& 0.0744 & 4.93	&  0.034  & 0.012  & 0.009	&  0.067  & 0.024  & 0.019 \\
0.060	& \!\!\!--\!\!\!& 0.100	& 0.0757 & 9.28	&  0.052  & 0.012  & 0.012	&  0.111  & 0.026  & 0.025 \\
0.060	& \!\!\!--\!\!\!& 0.100	& 0.0796 & 15.6	&  0.065  & 0.012  & 0.010	&  0.140  & 0.026  & 0.022 \\
\hline
0.100	& \!\!\!--\!\!\!& 0.150	& 0.119	& 6.99	&  0.058  & 0.017  & 0.014	&  0.072  & 0.022  & 0.017 \\
0.100	& \!\!\!--\!\!\!& 0.150	& 0.120	& 13.8	&  0.070  & 0.017  & 0.014	&  0.092  & 0.023  & 0.019 \\
0.100	& \!\!\!--\!\!\!& 0.150	& 0.124	& 24.2	&  0.148  & 0.017  & 0.019	&  0.191  & 0.023  & 0.025 \\
\hline
0.150	& \!\!\!--\!\!\!& 0.200	& 0.171	& 9.06	&  0.099  & 0.026  & 0.019	&  0.082  & 0.022  & 0.016 \\
0.150	& \!\!\!--\!\!\!& 0.200	& 0.171	& 19.2	&  0.119  & 0.026  & 0.021	&  0.101  & 0.022  & 0.018 \\
0.150	& \!\!\!--\!\!\!& 0.200	& 0.174	& 33.9	&  0.127  & 0.026  & 0.022	&  0.106  & 0.022  & 0.018 \\
\hline
0.200	& \!\!\!--\!\!\!& 0.250	& 0.221	& 11.2	&  0.150  & 0.037  & 0.028	&  0.087  & 0.022  & 0.017 \\
0.200	& \!\!\!--\!\!\!& 0.250	& 0.221	& 25.2	&  0.171  & 0.037  & 0.029	&  0.100  & 0.021  & 0.017 \\
0.200	& \!\!\!--\!\!\!& 0.250	& 0.224	& 43.5	&  0.151  & 0.037  & 0.032	&  0.085  & 0.021  & 0.018 \\
\hline
0.250	& \!\!\!--\!\!\!& 0.350	& 0.287	& 14.3	&  0.187  & 0.040  & 0.032	&  0.071  & 0.015  & 0.012 \\
0.250	& \!\!\!--\!\!\!& 0.350	& 0.288	& 33.4	&  0.187  & 0.040  & 0.032	&  0.068  & 0.015  & 0.012 \\
0.250	& \!\!\!--\!\!\!& 0.350	& 0.295	& 56.2	&  0.185  & 0.040  & 0.033	&  0.062  & 0.014  & 0.011 \\
\hline
0.350	& \!\!\!--\!\!\!& 0.500	& 0.400	& 20.0	&  0.396  & 0.065  & 0.056	&  0.070  & 0.012  & 0.010 \\
0.350	& \!\!\!--\!\!\!& 0.500	& 0.402	& 46.4	&  0.266  & 0.066  & 0.051	&  0.043  & 0.011  & 0.008 \\
0.350	& \!\!\!--\!\!\!& 0.500	& 0.411	& 74.1	&  0.288  & 0.063  & 0.050	&  0.041  & 0.009  & 0.007 \\
\hline
0.500	& \!\!\!--\!\!\!& 0.700	& 0.569	& 62.1	&  0.501  & 0.082  & 0.084	&  0.0204 & 0.0033 & 0.0035 \\
\hline
\end{tabular}
\end{table}

\end{document}